\documentstyle{mn}
\input{psfig}

\begin{document}

\title{New challenges for Adaptive Optics: Extremely Large Telescopes}

\author[M. Le Louarn et al.]{M.~Le~Louarn$^{1,2}$\thanks{e-mail: lelouarn@eso.org (MLL), nhubin@eso.org (NH), msarazin@eso.org (MS), atokovin@eso.org (AT)} N.~Hubin$^1$ M.~Sarazin$^1$ A.~Tokovinin$^1$\\
$^1$ European Southern Observatory, Karl Schwarzschild Str. 2 - D-85748 Garching, Germany\\
$^2$ CRAL - Observatoire de Lyon,  9, Av. Charles Andr\'e - F-69561 Saint Genis Laval, France\\
}

\date{Accepted 1999 November 3.
      Received 1999 November 4;
      in original form 1999 November 3}

\pagerange{\pageref{firstpage}--\pageref{lastpage}}
\pubyear{1994}

\maketitle

\label{firstpage}

   \begin{abstract} 

The performance of  an adaptive optics (AO) system on  a 100~m diameter ground
based telescope working in the visible range of the spectrum is computed using
an  analytical approach.   The target  Strehl ratio  of 60~\%  is  achieved at
0.5~$\mu$m with a  limiting magnitude of the AO  guide source near R$\sim$~10,
at the cost of an extremely  low sky coverage.  To alleviate this problem, the
concept of tomographic wavefront sensing in a wider field of view using either
natural guide stars  (NGS) or laser guide stars  (LGS) is investigated.  These
methods use  3 or 4  reference sources and  up to 3 deformable  mirrors, which
increase  up to  8-fold the  corrected  field size  (up to  60\arcsec ~at  0.5
$\mu$m).  Operation  with multiple NGS  is limited to  the infrared (in  the J
band  this approach yields  a sky  coverage of  50~\% with  a Strehl  ratio of
0.2).  The option  of  open-loop  wavefront correction  in  the visible  using
several bright NGS is discussed.  The LGS approach involves the use of a faint
(R $\sim$~22) NGS for low-order correction, which results in a sky coverage of
40~\% at the Galactic poles in the visible.  \end{abstract}

\begin{keywords}
Instrumentation: adaptive optics -- Atmospheric effects -- Telescopes
\end{keywords}

\section{Introduction}

The current generation of large ground based optical telescopes has a diameter
of the primary mirror in the 8 to 10 metre range.  Recently some thoughts have
been given to the next generation  optical telescopes on the ground.  In these
projects the diameter of the primary mirror lies in a range between 40 and 100
metres     (see    Gilmozzi~et~al.~1998\nocite{Gilmozzi98},    Andersen~et~al.
1999\nocite{Andersen99},   Mountain~1997\nocite{Mountain97}).    The  use   of
Adaptive Optics (AO, Roddier~1999\nocite{Roddier99}) in the visible is crucial
to obtain the full potential  in angular resolution, to avoid source confusion
for  extragalactic studies  at high  redshifts, and  to reduce  the background
contribution, dramatically increasing limiting  magnitude (the signal to noise
ratio is then proportional to  the square of telescope diameter).  Competition
with space  based observatories, providing  diffraction limited imaging  on an
8~m class  telescope (see  Stockman~1997\nocite{Stockman97}) is also  a driver
for AO correction in the visible with larger apertures.

In this  paper we address key  issues for a  visible light AO system  on these
Extremely Large Telescopes  (ELTs).  We have chosen  a telescope diameter
of 100~m, since it represents the  extreme case and we want to investigate the
limiting factors of AO on such a large aperture. We shall not address here the
astrophysical drivers  for such aperture  size, which are  presented elsewhere
(Gilmozzi~et~al.~1998\nocite{Gilmozzi98}).  We model the performance of an AO
system working  in the visible  on a 100~m  telescope, for an  on-axis natural
guide star (NGS) (section 2).  The sky coverage with this approach is close to
zero, because only  bright objects (R~$\sim$~10) can be  used as AO reference.
The use of a single artificial laser guide star (LGS) is ruled out by the huge
error   introduced    by   the    cone   effect   or    focus   anisoplanatism
(Foy~\&~Labeyrie~1985)\nocite{Foy85}.  We propose to use turbulence tomography
(i.e.    3D   mapping   of  turbulence,   Tallon~\&~Foy~1990\nocite{Tallon90},
hereafter    TF90)    combined    with   Multi-Conjugate    Adaptive    Optics
(Foy~\&~Labeyrie~1985\nocite{Foy85},  Beckers~1988\nocite{Beckers88}, hereafter
MCAO) as a way to increase the  fraction of the sky which can be observed.  In
section 3 we present the  main concepts involved in turbulence tomography.  In
section 4 we describe a fundamental  limitation of the corrected field of view
size corrected by a small (1-3)  number of deformable mirrors (DMs) and taking
into account real turbulence profiles.   A solution using 3 NGSs is presented,
where  the  correction  is  done  in  the  visible  (section  5)  and  in  the
near-infrared (section 6).  In section 7, another solution is presented, based
on 4  LGSs for visible correction.   In the following section,  we present and
quantify some  technical aspects of  AO on ELTs.   Finally, in section  9, the
conclusions are given.

\section{AO performance with an on-axis NGS}

There is a strong scientific interest  in visible light studies with the ELTs.
Using the software  described in Le~Louarn~et~al.~(1998)\nocite{Lelouarn98} to
perform analytical  calculations of  the AO system  performance, we  modeled a
system with a Strehl ratio (ratio of the peak intensity of the corrected image
to the peak  intensity of a diffraction limited image,  hereafter SR) of 60~\%
at   0.5~$\mu$m,   based   on   a  Shack-Hartmann   wavefront   sensor   (e.g.
Rousset~1994\nocite{Rousset94}).  The target SR is higher than required by the
scientific goals,  $\sim$~40~\%, to take into account  potential error sources
arising outside  the AO system (e.g.   aberration of the  optics or co-phasing
errors  of the telescope  primary mirror  segments).  Considering  the current
performance of AO systems, this is  a challenging goal.  However, the start of
the operation of ELTs is planned in 10-20 years from now, and AO technology is
bound  to  evolve  considerably.   The  atmospheric model  we  used  in  these
calculations corresponds to good  observing conditions at Very Large Telescope
observatory           of            Cerro-Paranal           in           Chile
(Le~Louarn~et~al.~1998)\nocite{Lelouarn98}.   The main  atmospheric parameters
and the AO hardware characteristics are summarized in Tab.~\ref{tab:ao_parm}.

The  effects  of  scintillation  on  the  wavefront  sensing  were  neglected.
Preliminary studies (Rousset 1999,  private communication) have shown that the
wavefront error contribution  could be between 20 and  30~nm rms, reducing the
SR by  $\sim$~10~\%.  The effects of  the outer scale of  turbulence were also
neglected.   Measurements (Martin~et~al.~1998\nocite{Martin98b})  yield values
usually between  20 to  30~m, significantly smaller  than the diameter  of the
ELT.   This is  a new  situation compared  to current  large  telescopes.  The
effect of the outer scale is mainly to reduce the relative contribution of low
order modes  of wavefront distortions  (Sasiela~1994)\nocite{Sasiela94} and to
decrease the  stroke needed  for the DM  to several microns,  independently of
telescope diameter.  This  relaxes constraints on the design  of DMs, but does
not change the overall on-axis system performance.

\begin{figure}
\centerline{\psfig{figure=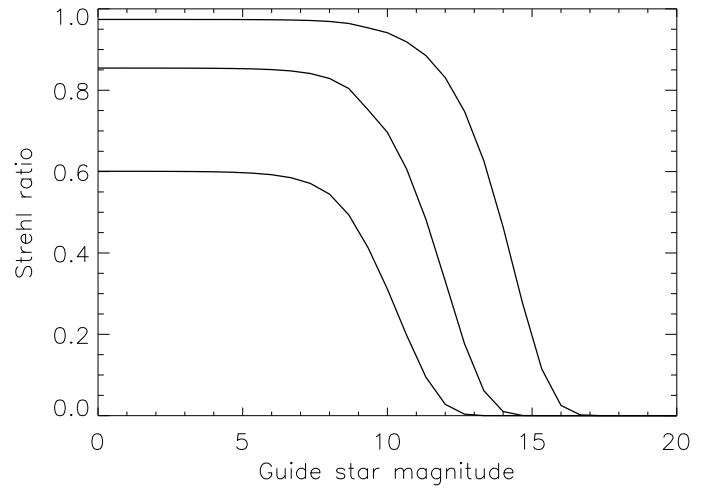,width=9cm}}
\caption{Strehl  ratio  versus magnitude   at 0.5, 1.25  and 2.2~$\mu$m  (from
bottom curve to  top)  for one on-axis NGS  with  the telescope pointing  at
zenith}
\label{fig:strehl_mag}
\end{figure}

The  simulation results are   presented  in Fig.~\ref{fig:strehl_mag}.   The
target SR of 60~\% is obtained in  the visible, providing 1.03 milli-arcsecond
(mas) diffraction limit  at 0.5~$\mu$m.  The  peak SR is  over 95~\% in K band
(2.2~$\mu$m), the diffraction limit being 4.5~mas.  Due  to the fine wavefront
sampling needed  for correction in  the visible,  the limiting  magnitude  (at
0.5~$\mu$m)  is $R \sim  10$, which is bright   compared to current AO systems
working   in        the    near-infrared  (around      $R     \sim   16$, e.g.
Graves~et~al.~1998\nocite{Graves98}).  This implies that with a single NGS the
sky coverage  is extremely small (see Rigaut~\&~Gendron~1992\nocite{Rigaut92},
Le~Louarn~et~al.~1998\nocite{Lelouarn98} for  a  more extensive  discussion on
sky coverage with AO systems).

To overcome this limitation, we propose  two different options, both involving
multiple reference  sources: NGS  and  LGS approaches  are investigated in the
following sections.

\begin{table}
\caption[]{AO simulation parameters.  Atmospheric values are given at
0.5~$\mu$m, WFS = wavefront sensor.}
\label{tab:ao_parm}
\begin{center}
\begin{tabular}%
{ll}
Telescope diameter & 100 m
\\
Number of actuators  & $\sim$~500000 
\\
WFS readout noise & 1 e$^-$
\\ 
WFS quantum efficiency & 90 \%
\\ 
WFS spectral bandwidth & 500 nm
\\ 
Transmission$^{1}$ & 40~\%
\\ 
WFS subaperture size & 16 cm
\\ 
Max. WFS sampling rate & 500 Hz
\\ 
Seeing & 0.5\arcsec
\\
Coherence time & 6 ms
\\
Isoplanatic angle & 3.5\arcsec
\\ 
\end{tabular}
\end{center}
$^{1}$Transmission  of the atmosphere and  telescope optics to the wavefront
sensor. For  visible  light  observations,  light must  be  split  between the
wavefront sensor path and imaging path.
\end{table}

\section{Turbulence tomography}

Turbulence  tomography is a  technique to  measure the  wavefront corrugations
produced by  discrete atmospheric  turbulent layers with  the help  of several
reference sources  (TF90).  Assuming  weak turbulence, the  phase corrugations
produced by each layer add linearly (Roddier~1981)\nocite{Roddier81}.  Knowing
the  configuration of the  guide sources  (position in  the sky,  height above
ground in the case  of an artificial star) and the altitudes  of the layers to
be measured, it is possible to reconstruct the phase at the selected turbulent
layers.  Foy~\&~Labeyrie~(1985)\nocite{Foy85} proposed  to use Multiple DMs to
correct them individually, a  concept called Multi-Conjugate AO (MCAO).  There
must be at  least as many measurements (number of guide  stars times number of
measurements points on  the pupil) as there are  unknowns (number of corrected
layers times  actuators on the  correcting mirrors).  Therefore, only  a small
number  (2-4) of  turbulent layers  can be  reconstructed, if  a  small number
($\sim$~4) of  reference sources are to  be used.  Recent  papers have tackled
the        problems        of        turbulence       tomography        (TF90,
Tallon~et~al.~1992\nocite{Tallon92},
Ragazzoni~et~al.~1999\nocite{Ragazzoni99},   Fusco~et~al.~1999\nocite{Fusco99},
Le~Louarn~\&~Tallon~2000\nocite{Lelouarn99}), and  reconstruction of turbulent
wavefronts has been demonstrated in numerical simulations.

The maximum size $\theta$ of the tomographic  corrected Field of View (FOV) is
given by geometrical considerations:
\begin{equation}
\label{eq:tomo_field}
\theta = \frac{D}{h_{max}}(1-\frac{h_{max}}{H}),
\end{equation}
where $D$ is  the diameter of the telescope, $h_{max}$ is  the height of the
highest turbulent layer  and $H$ the height of the guide  star (infinity for a
NGS).   As pointed  out by  TF90, in  circular geometry,  a small  fraction of
turbulence is not probed with this maximum FOV (pupil plane vignetting).  This
problem  can be  alleviated with  a modal  approach to  turbulence tomography,
which allows a slight interpolation  of the wavefront within the corrected FOV
(Fusco~et~al.~1999\nocite{Fusco99}). With a 100~m telescope it may be possible
to search reference stars  in a much larger patch of the  sky than with an 8~m
class  telescopes.   The  probability  to  find a  references  source  can  be
dramatically  increased  (Ragazzoni~1999\nocite{Ragazzoni99b}).   For a  100~m
telescope, the maximum tomographic field  is 17\arcmin~in diameter with a NGS,
or  13\arcmin~for LGSs,  if  the highest  turbulent  layer is  at 20~km  above
ground.

The   image is  corrected in the   whole  tomographic FOV  only  if the  whole
turbulence is concentrated in few  thin layers and  if each layer is optically
conjugated  to its  correcting mirror.  Taking   into account real  turbulence
profiles, we compute in the next  section the FOV  size which can be corrected
with few DMs and we show that it is much less than the tomographic FOV.

\section{Limitations of multi-conjugate AO}

\subsection{Turbulence vertical  profile measurements}

We     have    analyzed     the    PARSCA     (Paranal     Seeing    Campaign,
Fuchs~\&~Vernin~1993\nocite{Parsca}) balloon data on the vertical distribution
of  turbulence to  test the  assumption  that all  turbulence is  concentrated
within  a few  layers.  During  the site  testing campaign,  12  balloons were
launched at nighttime to measure the profile of the refraction index constant,
$C_n^2(h)$.      SCIDAR      (Scintillation     Detection     and     Ranging,
Azouit~\&~Vernin~1980\nocite{Azouit80})    measurements    were   also    made
simultaneously,         confirming        the         balloon        soundings
(Sarazin~1996)\nocite{Sarazin96}.

In   Tab.~\ref{tab:balloon_data} we summarize  some parameters  of the balloon
flights.  The  average Fried  parameter (Fried~1966\nocite{Fried66b}),  $r_0$,
was 19~cm at 0.5 ~$\mu$m, corresponding to a seeing of 0.55\arcsec -- slightly
better than the average seeing at Paranal, 0.65\arcsec.  Considering the small
time span during  which the balloons were launched  (19~days), these  data are
not fully representative of the site.  The  parameters have been corrected for
the height difference between the observatory (2638~m), and the launching site
(2500~m),  which  explains   the  slight difference  with   other publications
(e.g. Sarazin~1996\nocite{Sarazin96}).

\begin{table}
\caption[]{Balloon data for Cerro  Paranal.  The atmospheric coherence length,
$r_0$ and the  isoplanatic angle  $\theta_0$, are given   at a wavelength   of
0.5~$\mu$m.}
\label{tab:balloon_data}
\begin{center}
\begin{tabular}%
{lllll}
Flight  & Date & Time & $r_0$ (m) & $\theta_0$(\arcsec)
\\ 
38 & 10.03.92 & 3:30 & 0.32 & 3.80
\\ 
39 & 11.03.92 & 4:45 & 0.15 & 4.09
\\ 
40 & 12.03.92 & 1:30 & 0.21 & 3.58
\\ 
43 & 14.03.92 & 2:45 & 0.07 & 0.42
\\ 
45 & 15.03.92 & 1:00 & 0.19 & 2.05
\\ 
46 & 15.03.92 & 5:00 & 0.22 & 1.74
\\ 
48 & 16.03.92 & 4:10 & 0.17 & 2.48
\\ 
50 & 24.03.92 & 8:12 & 0.13 & 1.66
\\ 
51 & 25.03.92 & 2:43 & 0.23 & 2.03
\\ 
52 & 25.03.92 & 7:11 & 0.21 & 2.22
\\ 
54 & 23.03.92 & 4:10 & 0.22 & 2.81
\\ 
55 & 29.03.92 & 9:15 & 0.14 & 1.65
\\ 
\end{tabular}
\end{center}
\end{table}

\begin{figure*}
\centerline{\psfig{figure=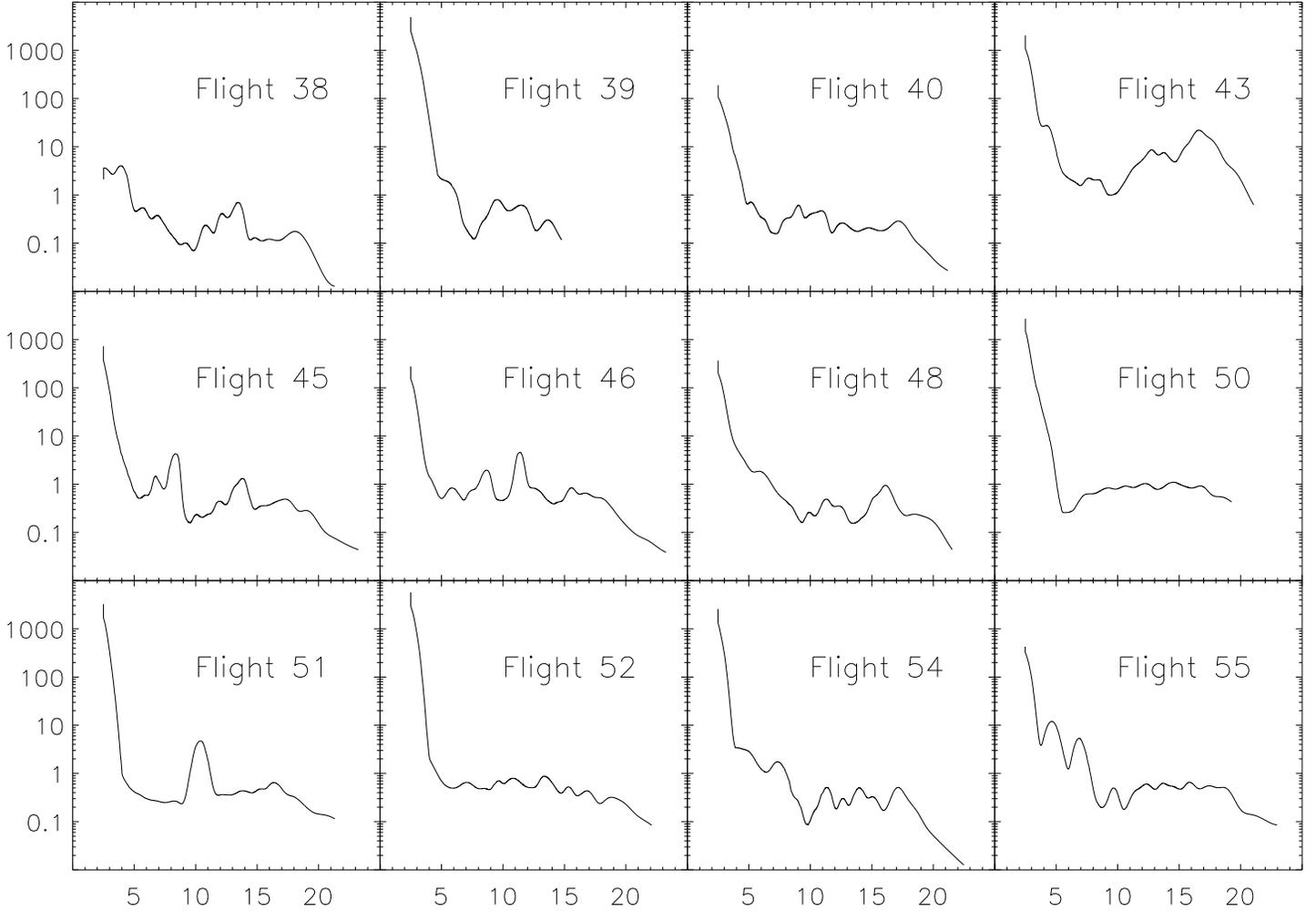,width=19cm}}
\caption{Profiles obtained by balloon  soundings above Cerro Paranal, smoothed
with a Gaussian  of standard deviation 500~m.  The  abscissae are altitudes in
kilometres  above sea  level.  Ordinates are  the  refractive index  structure
constant ($C_n^2$) in units of $10^{-17}$ m$^{-2/3}$.}
\label{fig:mean_balloon_data}
\end{figure*}

In  Fig.~\ref{fig:mean_balloon_data}  the  $C_n^2$ profiles  obtained by  the
balloon flights are plotted.  The  height resolution of the balloons is $\sim$
5~m.  For  clarity these measurements have  been convolved with  a Gaussian of
standard deviation 500~m.   The physics and formation of  very thin turbulence
laminae  is described in  Coulman~et~al.~1995\nocite{Coulman95}.  For  most of
the  flights the thin  turbulent layers  form larger  structures which  can be
identified  with the  turbulent layers  seen by  SCIDAR (see  for  example the
concentration of turbulence  near 15 km on flight  45, altitudes are expressed
in kilometres above sea level).  The strongest of these layers is the boundary
layer,  in the  first  kilometres of  the  atmosphere, present  on all  plots.
Another  layer, present  on most  flights,  is located  near 10-12~km.   These
measurements confirm the existence  of numerous layers.  However, a continuous
component of  small but significant amplitude  is also present on  most of the
soundings.

\subsection{Anisoplanatism in MCAO}

We  used the  high resolution  profiles (not  convolved  with a  Gaussian) and
applied       the       analytical         formula              derived     by
Tokovinin,~Le~Louarn,~\&~Sarazin~(2000)\nocite{Tokovinin99} to calculate   the
size  of the FOV $\theta_{M}$   which can be   corrected with $M$ deformable
mirrors.    This  is   a  generalized isoplanatic  angle    in  the  sense  of
Fried~(1982)\nocite{Fried82}, expressed as
\begin{eqnarray}
\label{eq:isoplanatism}
\theta_{M} &  =  & \left[  2.905 (  2 \pi / \lambda ) ^2  \right.  \nonumber \\ 
& & \left. \times \int C_n^2(h) F_{M}(h, H_1,
H_2, \cdots, H_{M}) {\rm d} h \right]^{-3/5}  , 
\end{eqnarray}
where $F_{M}$ is a function depending on the conjugation heights of the DMs,
$H_i$ the height  of conjugation, above ground.  This  expression assumes that
the  correction signals  applied  to each  DM  are optimized.   It assumes  an
infinite  turbulence outer  scale and  an  infinite $D/r_0$  ratio.  For  1~DM
conjugated to altitude $H_1$, Eq.~\ref{eq:isoplanatism} contains:
\begin{equation}
F_1(h)= | h - H_1 |^{5/3}, 
\end{equation}
which reduces to $F_1(h)=h^{5/3}$ if $H_1=0$ as in  conventional AO and yields
the classical $\theta_0$.  For a two mirror configuration the function has the
form:
\begin{eqnarray}
\lefteqn{F_2(h,H_1,H_2)  =  0.5 \left[ |h-H_1 |^{5/3} + |h-H_2 |^{5/3} \right. }\\
		& & - 0.5  |H_2-H_1|^{5/3}  \nonumber \\
               & &\left.   - 0.5 | H_2 - H_1 |^{-5/3} ( |h-H_1 |^{5/3} - |h-H_2 |^{5/3} )^2 \right] \nonumber
\end{eqnarray}

\begin{figure*}
\centerline{\psfig{figure=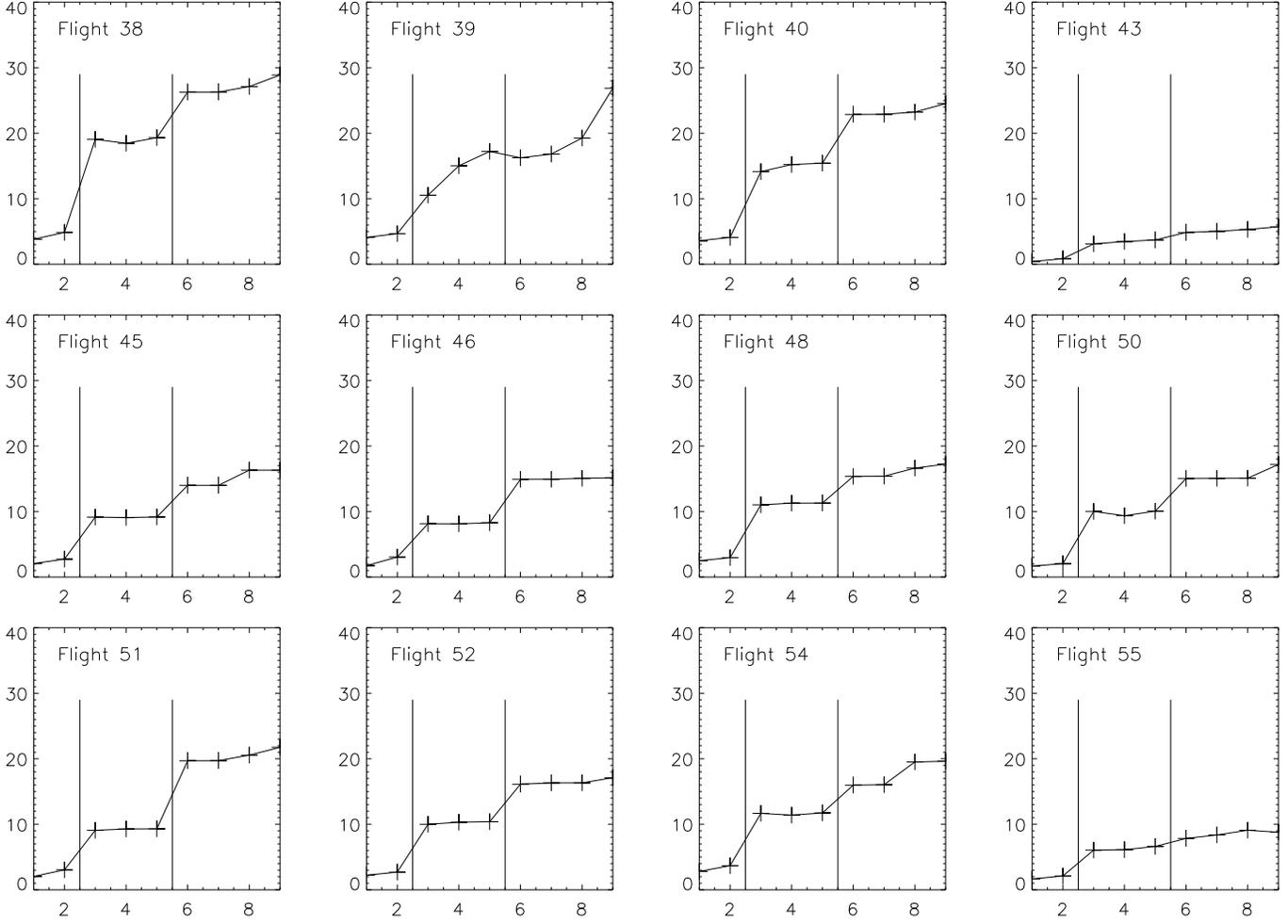,width=18cm}}
\caption{Isoplanatic angles  ($\theta_{M}$, ordinate  axis, in arcseconds at
0.5~$\mu$m) for different DM configurations (in abscissa, corresponding to $N$
in Tab.~3). Configurations 1, 2 are  for a single DM, 3
to 5 for 2 DMs and 6 to 9 for 3 DMs.}
\label{fig:balloon}
\end{figure*}

For 3 or more  DMs the expression  for $F_{M}$  is much  more  complex.  The
heights $H_i$ were computed  with a multi-parameter optimization algorithm  to
maximize $\theta_{M}$.  We explored the possibilities with 1, 2 and 3 DMs in
different altitude combinations, from all $H_i$  fixed to all $H_i$ optimized.
In the optimized   setups, the height of the   mirrors were adapted  for  each
flight to  maximize   the isoplanatic angle.  For  fixed   DMs,  we chose  the
conjugation height as the median of the heights  found by optimization. The DM
configurations are  summarized   in   Tab.~\ref{tab:mirror_summary},  and  our
results are shown in Fig.~\ref{fig:balloon}.

\begin{figure}
\centerline{\psfig{figure=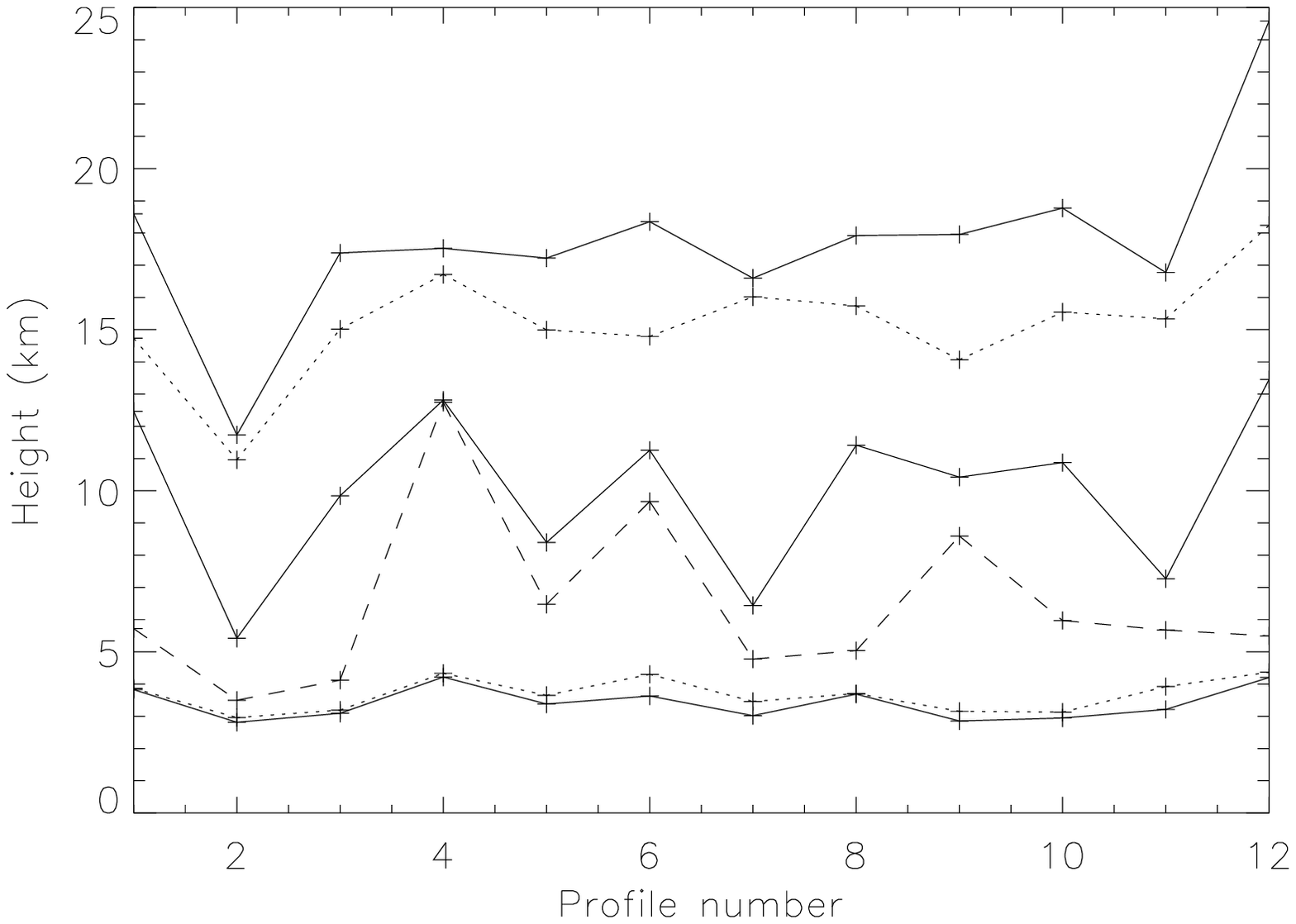,width=9cm}}
\caption{Optimized conjugate heights of the DMs  (in kilometres) as a function
of profile number.  Solid line is  for the 3 DM configuration,  dots for the 2
DM configuration and dash is for a single DM}
\label{fig:heights}
\end{figure}

With 3 DMs  the increase in $\theta_3$ (compared to  $\theta_0$) ranges from a
factor  of 2.6  to  13, depending  on  the profile.   The  median increase  of
$\theta_3$ is a  factor of 7.7, which means that the  isoplanatic angle in the
visible increases from 2.2\arcsec ~to 17\arcsec.  On particular nights (Flight
43  for  example,  which has  a  lot  of  extended high  altitude  turbulence)
$\theta_3$  stays small,  $\sim$~6\arcsec.  The  largest $\theta_3$  found was
28.9\arcsec  ~(Flight  38).   A  wavelength  of 2.2  $\mu$m  yields  a  median
$\theta_3$  of 102\arcsec  ~(for comparison,  $\theta_0$~=~13\arcsec).   A two
mirror configuration  brings improvement factors  between 1.6 and 8.7,  with a
median  of  4.6.  Therefore,  with  2  DMs, one  can  expect  to increase  the
isoplanatic angle to $\sim$~10\arcsec ~in the visible.  Adapting the conjugate
height of the DMs to profile  variations is not crucial ( $\theta_3$ increases
only by  $\sim$~7~\% when  using 3 optimized  heights instead of  fixed ones).
Fig.~\ref{fig:heights} shows  the optimal conjugate  heights for the DMs  as a
function  of  flight  number.   Three  main heights  are  identified:  ground,
$\sim$~10--12~km  and 15--20~km.   Considering the  observed stability  of the
optimum heights (due to the stability of the main turbulent layers), it is not
surprising that optimizing the heights does not improve significantly the FOV.
Notice  the   large  deviation  for  point   2  (Flight  39).    As  shown  in
Fig~\ref{fig:mean_balloon_data}, the turbulence was  located very low, and the
balloon reached only  a maximum altitude of $\sim$~15~km,  leaving part of the
turbulence unmeasured.

\begin{table*}
\caption[]{MCAO configurations and optimization results for Cerro-Paranal. The
columns    contain:    $N$   --    configuration    number    (same   as    in
Fig.~\ref{fig:balloon}), $M$ --  number of DMs, $H_1$, $H_2$,  $H_3$ -- median
conjugate  heights  of the  DMs  above  sea  level, $\theta_{M}$  --  median
isoplanatic angle in  arcseconds at 0.5~$\mu$m, $G$ --  gain in $\theta_{M}$
compared   to  the  median   $\theta_0$,  $G_{min}$   --  minimum   gain  in
$\theta_{M}$, $G_{max}$ -- maximum gain in $\theta_{M}$.}
\label{tab:mirror_summary}
\begin{center}
\begin{tabular}%
{lllllllll}
$N$ & $M$ & ${H_1}$, m & ${H_2}$, m & ${H_3}, m$ & ${\theta_{M}}$, $\arcsec$ & $G$ & $G_{min}$ & $G_{max}$
\\ 
1 & 1 & F: 2638 & - & - & 2.20 & 1
\\
2 & 1 & O: 5722 & - & - & 3.04 & 1.27 & 1.14 & 1.98
\\
3 & 2 & F: 3705 & F: 15337 & - & 10.00 & 4.50 & 1.39  & 8.55
\\
4 & 2 & F: 3381 & O: 15337 & - & 10.30 & 4.63 & 1.55 & 8.29
\\
5 & 2 & O: 3705 & O: 15337 & - & 10.38 & 4.66 & 1.66 & 8.68
\\
6 & 3 & F: 3381 & F: 10875 & F: 17922 & 15.96 & 7.17 & 2.17 & 11.80
\\
7 & 3 & F: 3381 & F: 10875 & O: 18030 & 16.04 & 7.20 & 2.25 &  11.80
\\
8 & 3 & F: 3381 & O: 11041 & O: 17842 & 16.61 & 7.46 & 2.38  & 12.18
\\
9 & 3 & O: 3381 & O: 10875 & O: 17922 & 17.23 & 7.74 & 2.56 & 12.97
\\ 
\end{tabular}

\medskip
O  --   optimized  altitude,  maximizing the  isoplanatic   angle\\
F  --   fixed altitude, taken to be the median of the optimized heights\\

\end{center}
\end{table*}

These results show that anisoplanatic effects occur in the visible even with 3
DMs used in  an MCAO approach.  They represent only one  site, on a relatively
short timescale.  Other sites with similar isoplanatic angles exist (e.g.  the
measurements  at   Maidanak,  Uzbekistan,  provide  a   median  $\theta_0$  of
2.48\arcsec (Ziad~et~al.~2000)\nocite{Ziad99}).   Moreover, the $\theta_{M}$
computed here is somewhat pessimistic,  since it contains a piston term (which
reduces the isoplanatic angle but does  not affect image quality) and does not
take into  account the  finite number of  corrected turbulent modes.   This is
similar to  the effect seen  with $\theta_0$, which  overestimates isoplanatic
effects (Chun~1998)\nocite{Chun98}.   Therefore, it is reasonable  to expect a
corrected FOV between  30\arcsec ~and 60\arcsec ~in diameter,  in the visible.
This is a considerable improvement over the few arcsecond isoplanatic field in
the visible (roughly equal to  $\theta_0$), but much less than the tomographic
FOV given by~Eq.~\ref{eq:tomo_field}.

We  suggest that the   site where an  ELT is  built be  optimized in  terms of
turbulence profiles, and not  only total turbulence, as it used to be in previous surveys.   

It is  more  effective to correct  a  few  strong layers  (even   if the total
turbulence  is higher),  than   a continuous repartition  of lower   amplitude
turbulence.  Indeed,  comparing for  example Flights~46 and   55 shows that  a
similar   $\theta_0  \sim$~1.7\arcsec   can be  well    corrected with  3  DMs
(Flight~46, $\theta_3 \sim$~14\arcsec) if turbulence  is concentrated in a few
peaks  (see    Fig.~\ref{fig:mean_balloon_data}),  whereas a  quasi-continuous
turbulence benefits much  less from   MCAO correction  (Flight~55,   $\theta_3
\sim$9~\arcsec).  The  location of the  turbulent layers  should  also be  as
stable  as possible,  to minimize  the  changes in  DM  conjugate height.   Of
course, some  other parameters of  the site will have  impact on the telescope
performance, like the wind (which is likely to be an important factor on such
a large structure).

\subsection{Required field of view}

\begin{figure}
\centerline{\psfig{figure=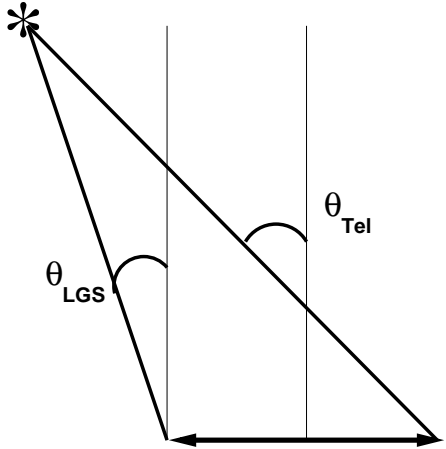,width=6cm}}
\caption{Required telescope field of view $\theta_{tel}$ compared to the
corrected field $\theta_{LGS}$ which corresponds to the positions of the
LGS. The need for a FOV much larger than $\theta_{LGS}$ is evident. }
\label{fig:lgs_angle}
\end{figure}

In tomographic wavefront  sensing using LGSs the  reference sources are placed
at the edges of the corrected field (TF90).  Therefore with 3 DMs the LGSs are
positioned  $\theta_{LGS}    = \theta_3$     apart.   The telescope     FOV,
$\theta_{tel}$, must  however be larger  (see Fig.~\ref{fig:lgs_angle}) for
the laser spots to be imaged by the telescope:
\begin{equation}
\label{eq:tel_fov}
\theta_{tel} = \theta_{LGS} + \frac{D}{H}.
\end{equation}
For a 100~m telescope  and a sodium LGS placed  at a 90~km height, and
for $\theta_{LGS}=$60\arcsec,  we get $\theta_{tel}  =$~290\arcsec,
or almost   5\arcmin ~in diameter. This   can   be a severe
requirement for the telescope optical design.

\section{Natural Guide Stars for visible correction}

The use of  several NGSs on an ELT  to increase the corrected FOV  and to find
reference   stars   outside   the    isoplanatic   patch   was   proposed   by
Ragazzoni~(1999)\nocite{Ragazzoni99b}.   He pointed  out that  with turbulence
tomography  the maximum  FOV which  can be  corrected increases  linearly with
telescope diameter, as shown  by Eq.~\ref{eq:tomo_field}.  Therefore, it would
be possible to use the huge  tomographic FOV to search for natural references.
This work assumed that anisoplanatism was not present in turbulence tomography
(turbulence concentrated in a few  thin layers).  In the previous paragraph we
have  shown that  this  is unfortunately  not  the case  with real  turbulence
profiles. As a consequence, if the  reference stars are much further away than
$\theta_{M}$  they  will not  benefit  from  AO  correction.  The  wavefront
measurement  would therefore be  done in  open-loop.  This  is a  very unusual
situation  in  AO (Roddier~1999)\nocite{Roddier99},  and  experiments must  be
carried out to verify the feasibility of that approach.

Moreover, our further studies show  that for widely separated NGSs, the errors
of tomographic  wavefront reconstruction with real turbulence  profiles can be
very  high.   So  the  use  of  3  NGSs in  a  wide  tomographic  field  seems
problematic.  Still,  we estimate  the sky coverage  for this option. 

Another constraint comes  from the telescope design.  The  telescope FOV of an
ELT  is a  strong  cost driver  and, at  the  moment, a  full tomographic  FOV
(17\arcmin) does not seem to be feasible.  Current optical designs for a 100~m
telescope (Dierickx~et~al.~1999)\nocite{Dierickx99}  provide a maximum  FOV of
12\arcmin.

\begin{figure}
\centerline{\psfig{figure=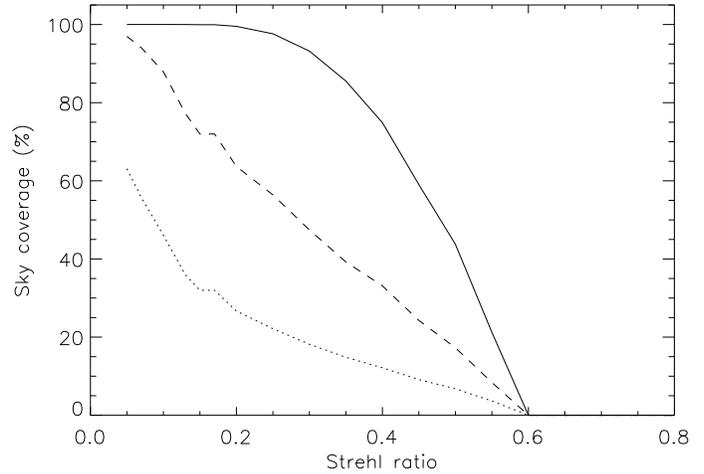,width=9cm}}
\caption{Sky  coverage   at 0.5~$\mu$m using  3 NGSs   in a  corrected  FOV of
12\arcmin~in diameter if wavefront sensing can be done  in open-loop. From top
to bottom curve: near Galactic plane, average latitude, near Galactic pole.}
\label{fig:sc_V_360}
\end{figure}

 We have computed the sky coverage  (SC) for the case when reference stars are
sought within a 12\arcmin ~FOV (Fig.~\ref{fig:sc_V_360}).  Full SC is obtained
only near the Galactic plane.  A 60~\% SC  can be achieved with a SR of 0.2 at
average Galactic coordinates ($l=180^\circ$,  $b=20^\circ$), or 30~\% near the
pole.

If a  telescope  design can be   improved to have the  maximum  FOV allowed by
tomography (Eq.~\ref{eq:tomo_field})  the SC  will be significantly increased.
A full SC can be achieved with a SR of 0.1 everywhere. SC of 50~\% is achieved
on the whole sky with a SR of at least 0.4.  Given  the performance of the AO
system shown in    Fig.~\ref{fig:strehl_mag},   the telescope FOV   size    is
identified here as a limiting factor for the sky coverage.

Initially we presumed  in these simulations that the  limiting magnitude for 3
NGSs    is   the    same   as    for   one    NGS,   e.g.     R    $\sim$   10
(Fig.~\ref{fig:strehl_mag}).  This is conservative with regards to the results
obtained  by   Johnston~\&~Welsh~(1994)\nocite{Johnston94}:  when  using  four
reference  stars, the  flux from  the  individual reference  sources could  be
divided  by four,  i.e.  a  gain  of 1.5  magnitudes. We  have therefore  also
studied  the  cases  where  the  limiting  NGS magnitudes  were  one  and  two
magnitudes  fainter.  Such gains  could be  achieved by  efficient tomographic
reconstruction algorithms.  If the limiting  magnitude can be increased by one
magnitude, a  SC of 40~\% at the  Galactic pole and 90~\%  at average Galactic
latitudes can be obtained with a  SR of 0.2. With the maximum tomographic FOV,
a SC of 50~\% is obtained with a SR 0.5 at the Galactic pole.

\section{Natural guide stars and correction in the infrared}

\begin{figure}
\centerline{\psfig{figure=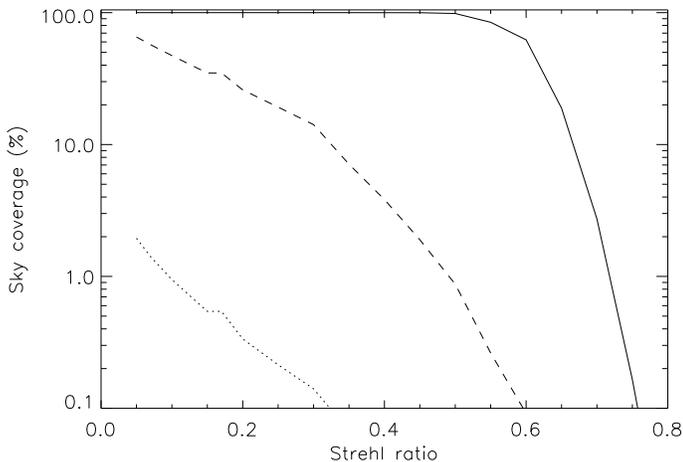,width=9cm}}
\caption{Sky  coverage in the  J band  using 3  NGSs with  a corrected  FOV of
3\arcmin~in diameter. From top to  bottom curves: near Galactic plane, average
latitude, near  Galactic pole.  Notice the logarithmic  scale of  the ordinate
axis.}
\label{fig:sc_J_180}
\end{figure}

The main  problem of  the NGS approach  in the  visible is caused  by residual
anisoplanatism.   This  problem is  alleviated  when  only  correction in  the
infrared is needed.  At 1.25~$\mu$m $\theta_{M}$ is increased by a factor of
3   compared   to    the   visible   (see   Eq.~\ref{eq:isoplanatism}).    For
$\sim$~60\arcsec ~FOV in the visible  (diameter), a 3\arcmin ~corrected FOV is
obtained.   The  limiting  magnitude, as  shown  by~Fig.~\ref{fig:strehl_mag},
increases from  R $\sim$~10 to  R $\sim$~13.  The  sky coverage is  plotted in
Fig.~\ref{fig:sc_J_180}.  It  shows that  with a Strehl  ratio of 0.2,  SCs of
0.4~\%, 30~\%, 100~\% are obtained  respectively at Galactic poles, at average
latitudes and in the Galactic disk.   If a 1$^m$ gain in limiting magnitude is
obtained compared to a single NGS, the coverages increase only slighlty.

At 2.2~$\mu$m, the  FOV is $\sim $~6\arcmin (diameter),  the limiting magnitude
is about  R $\sim$~15.   The sky coverage  is 10~\%  at the Galactic  pole and
complete elsewhere.

\section{Laser Guide Stars}

For astronomical AO systems  LGSs based on resonant  scattering in the  sodium
layer (Foy~\&~Labeyrie~1985)\nocite{Foy85} are usually considered because they
provide the highest reference source available, reducing the cone effect (also
called        focus       isoplanatism,    Foy~\&~Labeyrie~1985\nocite{Foy85},
Fried~\&~Belsher~1994\nocite{Fried94a},    Tyler~1994\nocite{Tyler94}).   This
effect is  due to the  finite altitude of  the laser guide  star.  It prevents
obtaining high AO correction in the visible already with 8~m telescopes.

\subsection{Power requirements}
\label{sec:power}
The laser power requirements for current  AO systems working in the near-IR is
of about 5~W  (Continuous-Wave, CW), providing LGS brightness  equivalent to a
$\sim     9^m$     guide    star     (Jacobsen~et~al.~1994\nocite{Jacobsen94};
Max~et~al.~1997\nocite{Max97};    Davies~et~al.~1998\nocite{Davies98}).    The
typical  sub-aperture  size  for  those  systems is  60~cm.   Scaling  to  the
subaperture size  in the  visible (16 cm)  to obtain similar  performance, the
power of the laser should be 14 times higher (assuming a linear scaling of the
guide star  brightness with  laser power), or  about 70~W (CW).   This scaling
does   not    take   saturation   of   the   sodium    layer   into   account.
Milonni~et~al.~1998\nocite{Milonni98}  provide an  analytical tool  to compute
the power  requirement in the case  of a pulsed  laser for a given  guide star
brightness with  saturation.  Using pulsed  laser characteristics of  the Keck
LGS implementation (Sandler~1999\nocite{Sandler99}) -- 11 kHz repetition rate,
100~ns pulse duration, -- we infer  that to receive the same number of photons
as for a  70~W~CW laser, a $\sim$~175~W pulsed laser  is needed. However,
considering Fig.~\ref{fig:strehl_mag}, we can  see that a 9$^{th}$ magnitude
guide star would provide a Strehl  ratio of 40~\%. Therefore, if a slight loss
of the AO system performance  is acceptable, a significantly smaller amount of
laser power would be sufficient.

One   could    instead   use   a   Rayleigh-scattering    based   LGS   system
(Fugate~et~al.~1994)\nocite{Fugate94}.  This  has the advantage  of being able
to  use any  laser  (producing a  bright  LGS at  an  arbitrary wavelength  is
currently not  a problem, see  Fugate~et~al.~1994\nocite{Fugate94}).  However,
the low altitude  of Rayleigh LGSs ($\sim$~15~km) reduces  its suitability for
tomography.  The position of the LGSs to obtain a zero corrected FOV (only the
cone effect is removed) is:
\begin{equation}
\label{eq:null_fov}
\theta_{null} = \frac{D}{H}.
\end{equation}
$\theta_{null} \sim$~23\arcmin~($D=$~100~m, $H=$~15~km), whereas the maximum
tomographic  FOV (Eq.~\ref{eq:tomo_field})  allowed by  the  highest turbulent
layer  (10~km,  optimistic  considering  Fig.~\ref{fig:mean_balloon_data})  is
$\sim$~11\arcmin  (for a  guide star  placed at  15~km).  Therefore,  the cone
effect can  not be fully corrected  with only 4  Rayleigh LGSs on ELTs  and we
will not consider this option in the remainder of this paper.

\subsection{Multiple sodium laser guide stars}

On a 100~m telescope the use of  a single LGS is totally impossible because of
the huge cone effect involved.  The  option of using multiple (4) sodium laser
guide  stars in  a tomographic  fashion has  therefore been  investigated.  We
should stress that  LGSs are placed on the edges of  the corrected FOV (TF90),
and therefore the problem of  open-loop wavefront measurements does not affect
this approach (the required FOV is given by $\theta_{M}$).  The problem with
LGSs in  turbulence tomography is that  the wavefront tilt  cannot be obtained
from  the LGS (Pilkington~1987)\nocite{Pilkington87}  and propagates  into the
global reconstructed wavefront.   In addition to global tilt,  other low order
modes (like forms of defocus and  astigmatism) have to be measured from an NGS
located             in            the             reconstructed            FOV
(Le~Louarn~\&~Tallon~2000)\nocite{Lelouarn99}.  Elaborate techniques have been
proposed    to    measure    the    tilt    from    the    LGS    (see    e.g.
Foy~et~al.~1995\nocite{Foy95},            Ragazzoni~1996\nocite{Ragazzoni96b}).
Unfortunately, real time correction has not been demonstrated.  If tilt can be
retrieved, this problem disappears and full SC is achieved.

To solve  the  problem of LGS  tilt  indetermination,  we  propose to  use  in
conjunction with   LGS a  very  low  order  wavefront sensor (for    example a
curvature  sensor, Roddier~et~al.~1988\nocite{Roddier88})  working on a  faint
NGS.  The limiting magnitude with 19  sub-apertures (4 sub-apertures across the
pupil) is currently of R~$\sim$~17 (Rigaut~et~al.~1998)\nocite{Rigaut98b} on a
3.6~m  telescope, with correction at  2.2~$\mu$m.  In Tab.~\ref{tab:gains}, we
summarize  the   scaling factors to  be  taken  into account   to convert this
limiting magnitude to that  of  a 100~m telescope with   a correction in   the
visible.  The  limiting magnitude is R~$\sim$~22.   This scaling is only valid
if compensation is done  in the visible, so   that wavefront sensing  benefits
from the  AO correction (Rousset~1994)\nocite{Rousset94}.  Otherwise, as shown
by Rigaut~\&~Gendron~(1992)\nocite{Rigaut92}, there   is no  gain  in limiting
magnitude for low order wavefront sensing on a  large aperture compared to 4~m
class telescopes.

\begin{table}
\caption[]{Scaling  of curvature   sensor  limiting  magnitude  from a   3.6~m
telescope  with  a   correction at  2.2~$\mu$m   to  a  100~m at   0.5~$\mu$m.
$D_{100}$  is the 100~m telescope diameter,  $D_{3.6}$ the 3.6~m diameter.
$r_0$    is given in the   visible   ($\sim$~0.2~m).  $\lambda_{05}$ is  the
correction wavelength of  the ELT, $\lambda_{2.2}$ the correction wavelength
of the 3.6~m telescope.  The factor 19 is  the number of sub-apertures on both
pupils.}
\label{tab:gains}
\begin{center}
\begin{tabular}%
{ll}
Factor & Flux gain (mag)
\\
Diameter $\propto (\frac{D_{100}}{D_{3.6}})^2$ & ~~+7
\\
Coherence time $\propto (\frac{\lambda_{2.2}}{\lambda_{0.5}})^{6/5}$ & ~-2 
\\ 
Measurement precision $\propto \frac{1}{19}(\frac{D_{100}}{r_0})^2$& ~+10
\\ 
Required precision $\propto (\frac{D_{3.6}}{D_{100}} \frac{\lambda_{0.5}}{\lambda_{2.2}})^2$ & -10
\\ 
Total &  ~~5
\\
\end{tabular}
\end{center}
\end{table}

We      used     a      model     of      the     Galaxy      developed     by
Robin~\&~Cr\'ez\'e~(1986)\nocite{Robin86}  to get  the probability  to  find a
star  of  a  given  magnitude  within  a given  FOV.   Considering  the  faint
magnitudes this  system will  be able to  use, we  also took into  account the
density  of   galaxies  in   the  sky.   We   used  galaxy  counts   given  by
Fynbo~et~al.~(1999)\nocite{Fynbo99},  based on  a combination  of measurements
from      the      Hubble      Deep      Fields     (North      and      South
(Williams~et~al.~1996)\nocite{Williams96})   and  the   ESO  NTT   deep  field
(Arnouts~et~al.~1999)\nocite{Arnouts99}).   Near  the  Galactic pole  galaxies
become more  numerous than stars for  magnitudes fainter than  $R \sim$~22.  A
bias may exist since not all of  these galaxies can be used as a reference due
to   their  size  (a   source  size   smaller  than   4~mas  was   assumed  in
Tab.~\ref{tab:gains}).  However, usually, the fainter the galaxies the smaller
they are.   We have assumed that  galaxies are distributed evenly  in the sky.
Poisson statistics give the probability to find a reference object for a given
AO limiting magnitude.

\begin{figure}
\centerline{\psfig{figure=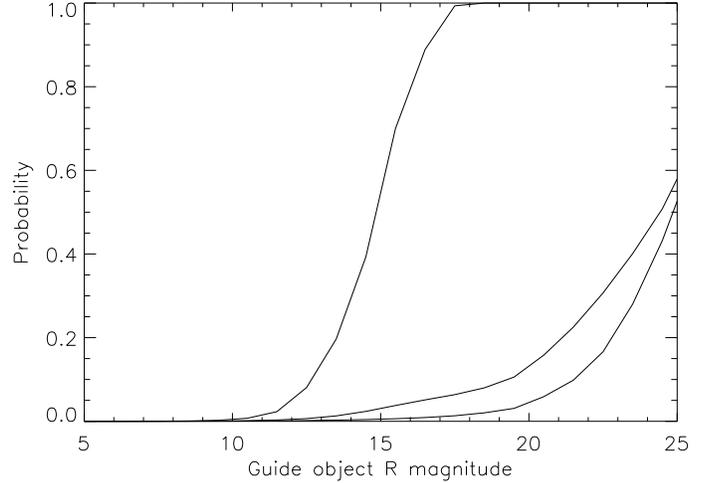,width=9cm}}
\caption{Sky   coverage with  4-LGS, $\theta_3  \sim$~30\arcsec,   top curve to
bottom: Galactic center, average position and Galactic pole.}
\label{fig:proba_lgs_15}
\end{figure}

\begin{figure}
\centerline{\psfig{figure=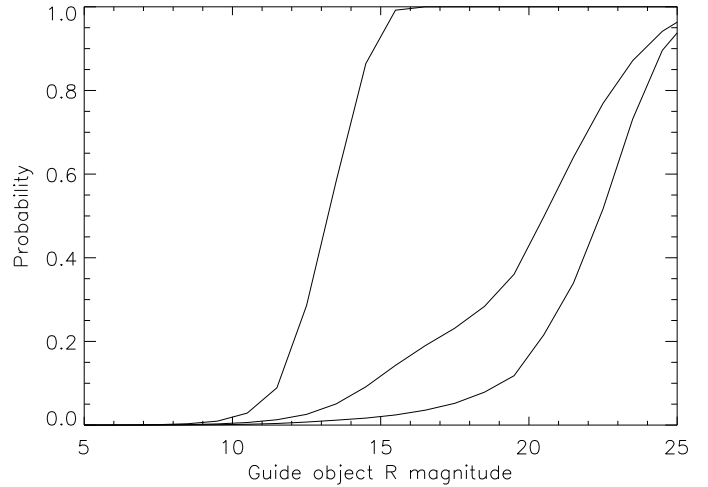,width=9cm}}
\caption{Sky coverage for the  4-LGS case, $\theta_{M}  \sim$~60\arcsec, top
curve to bottom: Galactic disk, average position and Galactic pole.}
\label{fig:proba_lgs_30}
\end{figure}

In Fig.~\ref{fig:proba_lgs_15} the probability  to find an NGS within a field
of 30\arcsec ~in  diameter is shown.  The SC is  $\sim$~13~\% for the Galactic
pole at R$\sim$22.  A twice  larger corrected isoplanatic angle (60\arcsec ~in
diameter, Fig.~\ref{fig:proba_lgs_30}), yields a  SC of 40~\% at  the poles,
70~\% at average latitudes and 100~\% near the Galactic plane.  Scaling the SR
vs limiting magnitude of current curvature systems, we expect a SR between 0.2
and 0.4 for this reference magnitude.   At a magnitude of $R \sim$~22, most of
the  wavefront references  sources will  be galaxies  when observing  near the
Galactic pole.

\section{Technical challenges}
In  the  previous  sections, we  have  shown  that  there are  no  fundamental
limitations  imposed by  the  laws  of atmospheric  turbulence  to building  a
visible light  AO system on  a 100~m optical  telescope.  In this  section, we
shall discuss the  technical difficulties which have to  be addressed to build
such a system.

\subsection{Wavefront sensor}

The number of sub-apertures of the  wavefront sensor impose the use of a large
detector.   Centroiding computations  require at  least 2$\times$2  pixels per
sub-aperture.  For  16~cm sub-apertures, this means that  the wavefront sensor
detector must  have at least 1250$^2$  pixels.  Moreover, if  guard pixels are
used,  this   number  could  increase  to  2500$^2$   (4$\times$4  pixels  per
sub-aperture).       The       pyramid      wavefront      sensor      concept
(Ragazzoni~\&~Farinato~1999\nocite{Ragazzoni99c}) requires only 2$\times$2 per
sampling  area and  therefore  could be  an  interesting alternative  to a  SH
sensor.  The  detector noise requirement  could be loosened slightly  from the
1$e^{-}$ level we have used, if  bright LGSs can be created in the atmosphere.
This is however  unlikely, since saturation problems in  the sodium layer will
arise (see section~\ref{sec:power}).

Currently, the  state of the art  detectors for wavefront  sensors are 128$^2$
(Laurent~et~al.~2000\nocite{Laurent00}).  The required  number of pixels could
however be reduced  by two means. One could use  a curvature wavefront sensing
method,  coupled  to a  CCD  detector.  This  approach  has  been proposed  by
Beletic,~Dorn~\&~Burke~(1999)\nocite{Beletic99}  and   has  the  advantage  to
reduce the  number of pixels needed  on the detector to  one per sub-aperture.
This would bring  the total required number of  pixels to $\sim$625$^2$, which
is realistic. It does not seem possible, with current technology, to produce a
bimorph  mirror   (usually  associated  to  curvature   sensors)  with  500000
actuators.  This problem  could be solved by coupling a  curvature sensor to a
piezo-stack deformable mirror but the approach clearly deserves more studies.

The read-out  rate of the  wavefront sensor detector  (SH) can be  obtained by
scaling  the typical  current  frame-rate in  the  IR ($\sim$  200~Hz) to  the
visible.   We obtain a  frame rate  of $\sim$~1.5~kHz.  Therefore, there  is a
choice to be made between a smaller number of pixels but high frame-rate and a
larger but slower system.

Both large number of pixels and high read-out speed can be achieved by butting
small  chips together,  with  multiple  read-out ports  (like  in the  Nasmyth
Adaptive           Optics          System           wavefront          sensor,
Laurent~et~al.~2000\nocite{Laurent00}), or  even more efficiently  by adapting
the         CCD        designing         technique         described        in
Beletic~et~al.~(1999)\nocite{Beletic99}   to  Shack-Hartmann   systems,  which
allows a  very efficient parallelization  of the read-out  process.  Therefore
the wavefront sensor  detector should not be technically  the most challenging
part of the AO system.

\subsection{Deformable mirror}

With  a typical  DM  diameter of  0.5~m which  could  be feasible  on a  100~m
telescope  (Dierickx~et~al.~1999\nocite{Dierickx99}), the  spacing requirement
between the DM actuators would be  0.8~mm. This a value ten times smaller than
on  existing DMs.  Therefore,  the production  of a  DM with  500000 actuators
clearly   requires  new   methods.    Current  development   based  on   MOEMS
(Micro-Opto-Electro-Mechanical systems) could lead to spacings down to 0.3~mm,
(e.g.                                      Bifano~et~al.~1997\nocite{Bifano97},
Vdovin~et~al.~1997\nocite{Vdovin97},   Roggeman~et~al.~1997\nocite{Roggeman97})
making possible a DM size of $\sim$20~cm.  One of the key issues in the design
of these DMs is the required stroke.  Assuming an outer scale of turbulence of
25~m and  a von~K{\'a}rm{\'a}n  model, a stroke  of $\pm 5  \mu$m (3~$\sigma$)
would be sufficient.   However, the actual turbulence spectrum  at low spatial
frequencies must be  measured on 8~m class telescopes  for realistic estimates
of the required stroke.

\subsection{Computing power}

By using Moore's law, which states  that the computing power doubles every 1.5
years,  the computing  power in  20 years  will be  increased by  a  factor of
10$^4$.

Current wavefront  computers have  a delay smaller  than 200 $\mu$s,  which is
compatible with use  in the visible (Rabaud~et~al.~2000\nocite{Rabaud00}). The
required computing  power increase can  therefore be estimated as  the squared
ratio of the number of controlled actuators:
\begin{equation}
\gamma = (\frac{N_{ELT}}{N_{IR-AO}})^2,
\end{equation}
where $N_{IR-AO}$ is the number of actuators of current IR AO systems (200),
and $N_{ELT}$  the number of  actuators for the  ELT (500000). We  get $\gamma
\sim 6  \times 10^{6}$.   However, this  does not take  into account  that the
cross-talk between  actuators will be  negligible for actuators far  away from
each other  and therefore  the interaction matrix  will be very  sparse.  This
will  reduce   significantly  the  computing  load.   If,   for  example,  the
interaction  matrix  (Boyer,  Michaud~\&~Rousset~1990\nocite{Boyer90}) can  be
broken  up into  6  times  $100\times100$ matrices,  the  likely evolution  in
technology would bring the adequate power in 20 years.

Another  possibility  would  be to  use  a  curvature  sensing, in  which  the
interaction  matrix is  almost diagonal  (if  no modal  control is  employed),
minimizing the computing power requirements.  However, this approach, as noted
earlier, seems to be prohibited by the availability of large bimorph mirrors.

\subsection{Optics}

The use of a small pitch between  the actuators of the DM allows to maintain a
small pupil  diameter: with a pitch  of 300~$\mu$m, the pupil  size is 187~mm.
This facilitates  the imaging of the  pupil on the  wavefront sensor detector.
Indeed,  with 625  sub-apertures across  the pupil,  the WFS  detector-size is
roughly 25~mm (assuming 2 pixels  per sub-aperture and 20 $\mu$m pixels).  The
reduction factor from  the pupil to the  detector is then 7.5, which  is not a
problem if each subaperture has a FOV of a few arcseconds.

Atmospheric dispersion  (AD) correction is  currently an unsolved  problem and
has to be tackled at the  level of telescope design.  For example, AD produces
an  elongation of  the object  of 184~mas  (if AD  is not  corrected, assuming
imaging between 0.5  and 0.6$\mu$m, at a zenith angle  of 30$^\circ$) which is
unacceptably high.  The design of  the AD corrector will be challenging, since
an  optimal combination  of glasses,  allowing a  correction with  an accuracy
better than 1~mas must be found.  The physical sizes of these AD correctors is
also  a problem,  because  of the  large  size of  the  optics.  The  required
precision puts severe constraints on the measurement of atmospheric parameters
(air temperature, humidity, pressure).

For the  multi-NGS scheme this  problem is even  more crucial, since  the NGSs
must be  far apart  to increase  the sky coverage,  and will  therefore suffer
immensely from AD.  The multi-LGS has  the advantage to be in sensitive to AD,
because the sources are highly monochromatic.

If proper  correctors cannot be  built for technological reasons,  narrow band
operation of the telescope should be used if the highest spatial resolution is
required: at 30$^\circ$ from zenith a bandpass of 0.4~nm produces a dispersion
of  $\sim$1~mas, if  no correction  is made.   The use  of 3D  detectors (like
integral  field  spectrographs), would  solve  the  problem,  since images  in
different colors can then be disentangled.

In the multi-NGS case, if anisoplanatism limits the correction and the sources
do not  benefit from  AO correction, non-common  path aberrations  between the
sources will be difficult to maintain.

\subsection{Laser spot elongation}

The   atmospheric  sodium  layer   is  roughly   10~km  thick   (e.g.   Papen,
Gardner~\&~Yu~1996\nocite{Papen96}).  This causes the  LGS to be extended, for
sub-apertures which are  not on the optical axis of  the telescope (assuming a
projection of the LGS from behind  the secondary mirror of the telescope). The
apparent size of a laser spot is given by simple geometry:

\begin{equation}
\theta_{spot} \sim \frac{\Delta H d }{H_{Na}^2}
\end{equation}
where $\Delta H$ is the thickness  of the sodium layer (10~km), $H_{Na}$ the
altitude of the  Sodium layer ($\approx$ 90 km), $d$ is  the separation of the
beam-projector  and  the  considered   sub-aperture.   With  $d$=50~m  we  get
$\theta_{spot} \sim  13$\arcsec. Since the  multiple LGSs will  be off-axis,
the spots will be even more elongated.  This is clearly too large for standard
wavefront sensors, which typically have a field of view of 2-3\arcsec. Several
methods  have been proposed  to eliminate  spot elongation.   The conceptually
simplest is to  use a pulsed laser and  to select only a small  portion of the
laser stripe  by time  gating the photons  coming from  the LGS. This  has the
advantage of being technically simple, at the cost of the effective brightness
of  the  LGS.  Other  solution  have been  proposed  in  the literature  (e.g.
Beckers~(1992)\nocite{Beckers92b}, improving  the previous scheme  by shifting
the wavefront  sensor measurements synchronously  with the propagation  of the
beam in the  sodium layer, thus removing  the loss of photons at  the price of
complexity. Other  less technically challenging solutions  should certainly be
investigated.

\section{Conclusions}

Although  a realization of  an adaptive  optical system  working with  a 100~m
telescope in  the visible represents a  technical challenge, it  is shown here
that  very large  aperture opens  a  number of  new possibilities  and such  a
correction becomes  feasible for a significant  fraction of the  sky.  The new
approaches involve either use of several widely spaced bright NGS (in the near
IR)  or  a very  faint  NGS  combined  with few  LGS.   In  both cases  a  3-D
tomographic measurement  of instantaneous  phase screens is  needed. Wavefront
correction will be made with few  (2-3) DMs conjugated to the optimum heights;
in this way the FOV size is increased $\sim$~8 times compared to the single-DM
AO systems, and  FOV diameter may reach 1\arcmin  ~in the visible.  Additional
criteria for site selection related to operation in this mode are formulated.
\begin{table}
\caption[]{Summary of  the studied systems. The  system NGS (1) is  based on 3
NGS  with a  wide FOV  (6\arcmin)  to search  for guide  stars. The  wavefront
sensing is done  in open loop.  NGS (2)  is also a 3 NGS  system, but optmized
for  the near  IR.  The LGS  stands for  the  4-LGS system  optimized for  the
visible.}
\label{tab:summary}
\begin{center}
\begin{tabular}%
{llll}
  & NGS (1) & NGS  (2) & LGS 
\\
Wavelength ($\mu$m) & 0.5 & 1.25  & 0.5 
\\
Strehl ratio (peak) & 0.6  & 0.85 & 0.6
\\
Max corrected FOV$^{(1)}$ & $\sim$60\arcsec & 3\arcmin & 60\arcsec
\\ 
Sky Coverage$^{(2)}$ & 60\% & 30\% & 70\%
\\
Number of NGS & 3 & 3 & 1
\\ 
Number of LGS & 0 & 0 & 4
\\ 
Technical difficulty & very high & low & medium
\\ 
\end{tabular}
\end{center}
$^{(1)}$: Limited by residual anisoplanatism\\
$^{(2)}$: with a Strehl ratio of 0.2, average Galatic latitude\\
\end{table}

\section*{acknowledgements}

The authors would like to thank Roberto Ragazzoni for many useful discussions,
Johan  Fynbo for his  data on  the magnitude  distribution of  galaxies.  This
paper  benefitted  from many  discussions  with  B.~Delabre, Ph.~Dierickx  and
R.~Gilmozzi regarding the design of  100~m telescopes. We are also grateful to
an anonymous referee  for improving the quality of this  paper.  This work was
done with the help of the  European TMR network ``Laser guide star for 8-metre
class telescopes'' of the European Union, contract \#ERBFMRXCT960094.


\bsp
\label{lastpage}

\end{document}